\documentclass[journal=jacsat,manuscript=article
]{achemso}

\usepackage{chemformula}
\usepackage[T1]{fontenc}
\usepackage{graphicx}% Include figure files
\usepackage{dcolumn}% Align table columns on decimal point
\usepackage{bm}
\usepackage{color}
\usepackage{amsmath}
\usepackage{txfonts}
\usepackage{upgreek}
\usepackage{romannum}
\usepackage{indentfirst}

\AtBeginDocument{\pagenumbering{arabic}}

\title{Realizing a Superconducting Square-Lattice Bismuth Monolayer}

\author{Eunseok Oh$^{\parallel}$}
\affiliation [1]
{Center for Artificial Low Dimensional Electronic Systems, Institute for Basic Science (IBS), Pohang 37673, Korea}
\alsoaffiliation [2]
{Department of Physics, Pohang University of Science and Technology, Pohang 37673, Korea}
\author{Kyung-Hwan Jin$^{\parallel}$}
\affiliation [1]
{Center for Artificial Low Dimensional Electronic Systems, Institute for Basic Science (IBS), Pohang 37673, Korea}
\author{Han Woong Yeom}
\email{yeom@postech.ac.kr}
\affiliation [1]
{Center for Artificial Low Dimensional Electronic Systems, Institute for Basic Science (IBS), Pohang 37673, Korea}
\alsoaffiliation [2]
{Department of Physics, Pohang University of Science and Technology, Pohang 37673, Korea}

\begin{document}

\maketitle

%TOC
%%%%%%%%%%%%%%%%%%%%%%%%%%%%%%%%%%%%%%%%%%%%%%%%%%%%%%%%%%%%%%%%%%%%%%%%%%%%%%%%%%%%%%%%
%%%%%%%%%%%%%%%%%%%%%%%%%%%%%%%%%%%%%%%%%%%%%%%%%%%%%%%%%%%%%%%%%%%%%%%%%%%%%%%%%%%%%%%%
%%%%%%%%%%%%%%%%%%%%%%%%%%%%%%%%%%%%%%%%%%%%%%%%%%%%%%%%%%%%%%%%%%%%%%%%%%%%%%%%%%%%%%%%

\begin{tocentry}
\centering
\includegraphics[width=82.5mm]{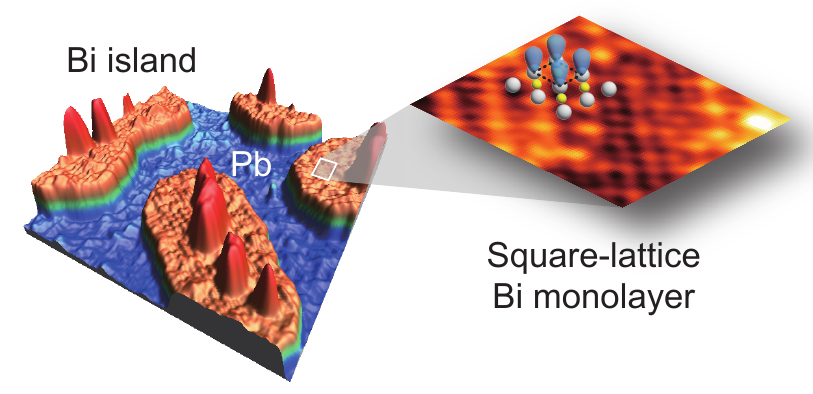}
\end{tocentry}

%Abstract
%%%%%%%%%%%%%%%%%%%%%%%%%%%%%%%%%%%%%%%%%%%%%%%%%%%%%%%%%%%%%%%%%%%%%%%%%%%%%%%%%%%%%%%%
%%%%%%%%%%%%%%%%%%%%%%%%%%%%%%%%%%%%%%%%%%%%%%%%%%%%%%%%%%%%%%%%%%%%%%%%%%%%%%%%%%%%%%%%
%%%%%%%%%%%%%%%%%%%%%%%%%%%%%%%%%%%%%%%%%%%%%%%%%%%%%%%%%%%%%%%%%%%%%%%%%%%%%%%%%%%%%%%%

\begin{abstract}
Interplay of crystal symmetry, strong spin$-$orbit coupling (SOC), and many-body interactions in low dimensional materials provides a fertile ground for the discovery of unconventional electronic and magnetic properties and versatile functionalities. 
Two-dimensional (2D) allotropes of group 15 elements are appealing due to their structures and controllability over symmetries and topology under strong SOC.
 Here, we report the heteroepitaxial growth of a proximity-induced superconducting 2D square-lattice bismuth monolayer on superconducting Pb films.
The square lattice of monolayer bismuth films in a $C_4$ symmetry together with a stripey moir\'e structure is clearly resolved by our scanning tunneling microscopy and its atomic structure is revealed by density functional theory (DFT) calculations.
A Rashba-type spin-split Dirac band is predicted by DFT calculations to exist at the Fermi level and becomes superconducting through the proximity effect from the Pb substrate.
We suggest the possibility of a topological superconducting state in this system with magnetic dopants/field.
This work introduces an intriguing material platform with 2D Dirac bands, strong SOC, topological superconductivity, and the moir\'e superstructure. 
\end{abstract}


%Introduction
%%%%%%%%%%%%%%%%%%%%%%%%%%%%%%%%%%%%%%%%%%%%%%%%%%%%%%%%%%%%%%%%%%%%%%%%%%%%%%%%%%%%%%%%
%%%%%%%%%%%%%%%%%%%%%%%%%%%%%%%%%%%%%%%%%%%%%%%%%%%%%%%%%%%%%%%%%%%%%%%%%%%%%%%%%%%%%%%%
%%%%%%%%%%%%%%%%%%%%%%%%%%%%%%%%%%%%%%%%%%%%%%%%%%%%%%%%%%%%%%%%%%%%%%%%%%%%%%%%%%%%%%%%

Low dimensional systems with various symmetries under strong spin-orbit coupling (SOC) have been widely investigated to achieve exotic electronic and spintronic properties and functionalities including topological properties, helical spin textures, and Dirac fermions~\cite{Bhimanapati2015,Liu2019,Kou2017,Ahn2020}
These research activities have been extended to artificial two-dimensional (2D) heterostructures by combining materials of strong SOC with those of superconducting and ferromagnetic properties, which provide a route to Majorana quasiparticles~\cite{Jack2021} and innovative quantum devices~\cite{Sierra2021}.
In particular, elemental 2D materials composed of heavy group 15 elements (As, Sb, Bi) have been highlighted because of their strong SOC and the existence of various 2D allotropes such as $\alpha$, $\beta$, $\gamma$, $\zeta$ forms~\cite{Zhang2016,Zhang2018,Wang2015,Markl2018,Hirahara2011,Li2022}.
Different types of allotropes were obtained on different substrates~\cite{Markl2018,Hirahara2011,Li2022} and further controlled through annealing~\cite{Bian2009} and changing thickness~\cite{Scott2005, Nagao2004} based on subtle energetics of different allotropes, which provides wide phase space for physics and high degree of freedom in material design for devices. 

A prototype example is monolayers of bismuth with the strongest SOC.  
Mainly two different 2D allotropes of Bi monolayers have been investigated, namely, rectangular ($\alpha$-Bi) and hexagonal ($\beta$-Bi) allotropes.
Initially, many studies focused on the hexagonal allotrope ($\beta$-Bi) which is related to the natural cleavage plane (111) of the bulk bismuth crystal~\cite{hoffman2006}. 
This system attracted huge interest as a candidate of 2D topological insulator (TI) to realize a quantum spin Hall edge state~\cite{Wada2011,Yeom2016,Murakami2006,Liu2011}. 
Few layer Bi(111) films were grown on various substrates~\cite{Drozdov2014,Yang2012,Hirahara2011,Chen2012,Yao2016,Sun2017,Kim2014,Peng2018}, and its flat layer version, bismuthene, was also realized with a large band gap and a well defined quantum spin Hall edge state ~\cite{Reis2017,Sun2022}. 
A puckered structure of $\alpha$-Bi has also been observed widely, which has a black phosphorous (BP)-like (A17) structure with nonsymmorphic symmetry~\cite{Nagao2004,Li2014,Qiao2014,Jin2019}.
Note that this puckered structure is intrinsically composed of two atomic layers paired. 
While this structure is topologically trivial, inversion symmetry breaking can induce an in-plane polarization field to result in ferroelectricity and substantial Berry curvature dipole~\cite{Xiao2018,Jin2021}.
Moreover, the strain could readily drive it into a topological insulator phase by tuning the buckling height~\cite{Lu2015,Li2017}. 
The other structure, a monolayer of Bi(110) bulklike (A7) structure (Figure 1a) has the nonsymmorphic and glide mirror symmetry and is known to induce 2D Dirac fermions~\cite{Bian2014,Wieder2016,Robin2017}.
Despite rich electronic and topological properties of (110) films, the single atomic layer of rectangular Bi is rarely realized~\cite{Kowalczyk2017} due probably to its instability with unsaturated p$_z$ dangling bonds~\cite{hoffman2006,Bian2014}; it was reported only as a wetting layer or a sandwiched layer in very thin Bi(110) films~\cite{Kowalczyk2011,Kowalczyk2013,Nagase2018,Wang2022}.

In this work, we successfully fabricate a square-lattice monolayer of bismuth on Pb(111) thin films grown on a Si(111) substrate by low temperature deposition. 
STM images reveal that the unitcell of a rectangular bismuth monolayer has equivalent longitudinal and vertical unit length of 4.7~\r{A}, being very close to that of bulk Bi(110).
It exhibits a characteristic stripy moir\'e superstructure, whose atomic structure is successfully analyzed in STM and DFT calculations. 
DFT calculations further show that strongly spin-split Dirac band is induced at the Fermi level by Rashba effect due to the strong SOC with sublattice symmetry breaking. 
The proximity-induced superconducting gap is confirmed by scanning tunneling spectroscopy measurements at 4.3~K, establishing a superconducting spin-split Dirac fermion system. 
We suggest that this system can host a topological superconducting (TSC) phase by the absorption of magnetic impurities or an uniform magnetic field.

%Result
%%%%%%%%%%%%%%%%%%%%%%%%%%%%%%%%%%%%%%%%%%%%%%%%%%%%%%%%%%%%%%%%%%%%%%%%%%%%%%%%%%%%%%%%
%%%%%%%%%%%%%%%%%%%%%%%%%%%%%%%%%%%%%%%%%%%%%%%%%%%%%%%%%%%%%%%%%%%%%%%%%%%%%%%%%%%%%%%%
%%%%%%%%%%%%%%%%%%%%%%%%%%%%%%%%%%%%%%%%%%%%%%%%%%%%%%%%%%%%%%%%%%%%%%%%%%%%%%%%%%%%%%%%

\section{RESULTS AND DISCUSSION}

\begin{spacing}{1.12} 

\begin{figure*}[ht!]
\centering
\includegraphics[width=164.6mm]{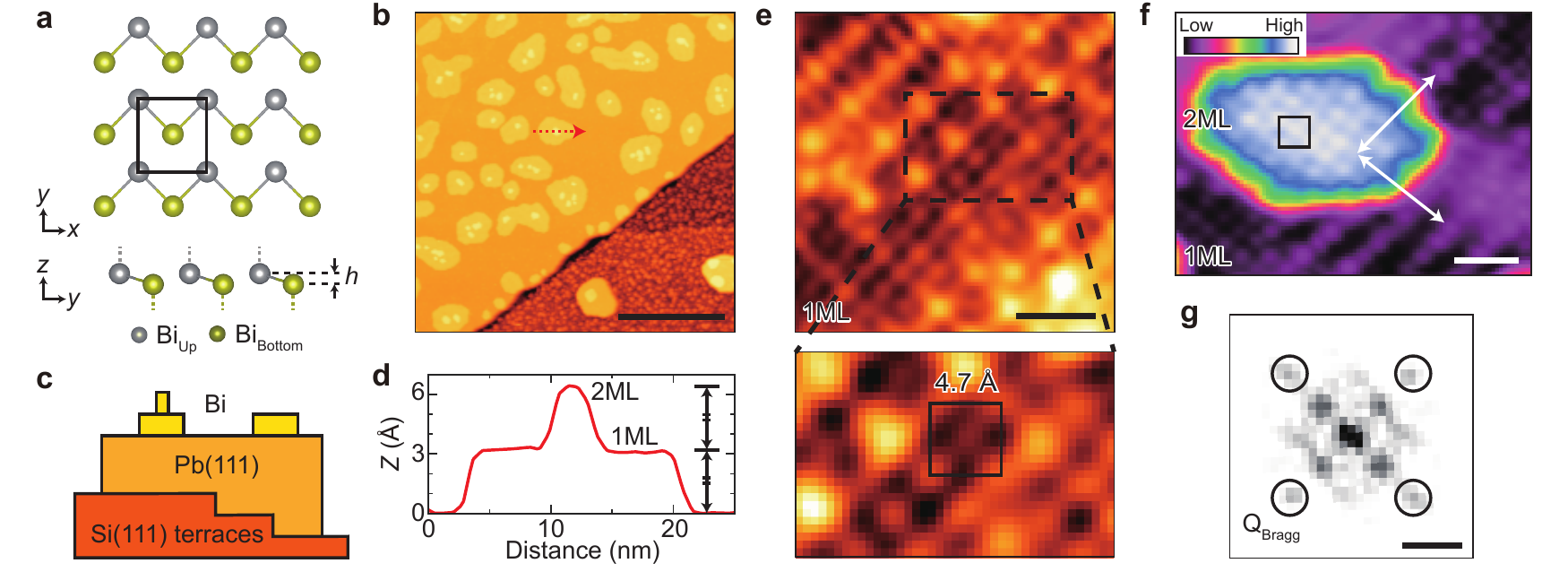}
\caption{\label{fig1} \textbf{Growth of a square-lattice bismuth monolayer. (a) Schematic model of a rectangular Bi(110) monolayer. The rectangle denotes a unit cell and the finite buckling height ($h >$ 0) is considered.
(b) STM image of a Bi/Pb/Si(111) system (150 $\times$ 150~nm$^2$, sample bias $V_{\textrm{bias}}$ = $-$1~V, tunneling current $I_\textrm{t}$ = 20~pA) and its (c) cross-sectional schematics. Scale bar, 50~nm.
(d) Line profile crossing a Bi island on a Pb(111) film [along the red dotted arrow in (b)]. 
Atomically-resolved STM images of (e) a monolayer (4 $\times$ 4~nm$^2$, $V_{\textrm{bias}}$ = $-$500~mV, $I_\textrm{t}$ = 100~pA, zoom-in image for the regions of dashed rectangle in the bottom) and (f) a bilayer (5.5 $\times$ 4~nm$^2$, $V_{\textrm{bias}}$ = $-$115~mV, $I_\textrm{t}$ = 100~pA) Bi island. Scale bars, 1~nm. 
Squares denote the $1\times1$ unit cell and arrows in (f) follow atomic rows of the Bi monolayer. 
(g) Fourier transformation of the STM image  of a Bi monolayer in (e). Bragg peaks (Q$_{\textrm{Bragg}}$) are denoted by circle. Scale bar, 1.3~\r{A}$^{-1}$.
}}
\end{figure*}

\end{spacing}

\subsection{A square-lattice bismuth monolayer on Pb films}
Figure 1b shows a large scale STM topographic image after the low temperature deposition of Bi on the Pb/Si(111) system. 
As is well known, Pb films form large flat-terrace islands over a few step edges of the Si(111) substrate reaching to a width of few hundred nm (see the schematics in Figure 1c)~\cite{Pan2011}. 
The top left part of this image covers a part of such a Pb film, which has a thickness of 8-10 monolayers (ML) with the monolayer height $h_{\textrm{Pb}}$ measured as 2.85~\r{A}. 
The deposition of Bi leads to the formation of small islands of few ten nm on top of the flat Pb terrace.   
These extra islands are not observed when the substrate is held at room temperature (see Supporting Information Part \Romannum{1} for details and Figures S1 and S2).
One can also see that most of extra islands are single layers but have second layers with much smaller areas. 
The height profile of one representative island is shown in Figure 1d, where one can measure the diameter of the 1~ML (2~ML) island as 16~nm (3~nm) and the layer height as $\sim$3.2~\r{A} for both layers.
The monolayer height is distinct from that of Pb islands and half of those of most of $\alpha$-Bi films reported, which correspond to the puckered structure~\cite{Nagao2004,Kowalczyk2011,Wang2022}.
The growth morphology is clearly contrasted from the formation of Pb-Bi alloy films through the codeposition of Pb and Bi~\cite{Mustafa2007,Ming2021}.
This clearly indicates that the extra islands are Bi islands in a distinct \textit{monolayer} structure from that of the $\alpha$-Bi films reported previously.

Figure 1e,f show the atomically-resolved STM images on 1 and 2~ML Bi islands. 
The square lattice is unambiguously observed, whose longitudinal and vertical lattice constants are the same as 4.7~\r{A}.
This is indeed very close to that of $\alpha$-Bi reported (4.5~\r{A} $\times$ 4.7~\r{A})~\cite{Nagao2004,Lu2015,Peng2019}.
However, the centered Bi atom in a unitcell locates at the symmetric position to secure two mirror planes in contrast to Bi(110). 
This structure, thus, ensures a C$_4$ symmetry to correspond to a \textit{square} lattice. 
The 2D fast Fourier transform (FFT) of topographic image of 1 ML (Figure 1g) confirms the square lattice.
A similar structure was previously observed as a surface termination layer of puckered $\alpha$-Bi~\cite{Kowalczyk2017}.
However, this surface layer is substantially compressed with a lattice of 3.9~\r{A} $\times$ 3.9~\r{A} and has a buckling of 1.5~\r{A}, which is similar to a half of the puckering height in $\alpha$-Bi of 3.3~\r{A}.
In clear contrast, the present square-lattice monolayer has only a very small buckling of 0.1-0.2~\r{A} (Figure 2e). 
The C$_4$ symmetric square lattice is shared by 1 and 2 ML islands and the position of Bi atoms in the second layer is matched with that of the first layer (sitting on on-top sites) (Figure 1f and see extended image in Figure S3).
This tells the formation of one bilayer Bi(110)-like structure through vertical Bi-Bi bondings between layers.
This layer stacking guarantees the presence of $p_z$ dangling bonds in the monolayer, which is the main characteristic of a single layer of Bi(110).
This is in clear contrast with the previous report of similar Bi(110) layers, which was suggested to exist as an interlayer sandwiched by puckered Bi layers~\cite{Nagase2018}.
We thus conclude that a monolayer form of a square-lattice bismuth is distinct to the present work.

\begin{spacing}{1.12}
\begin{figure*}[htb!]
\centering
\includegraphics[width=84.6mm]{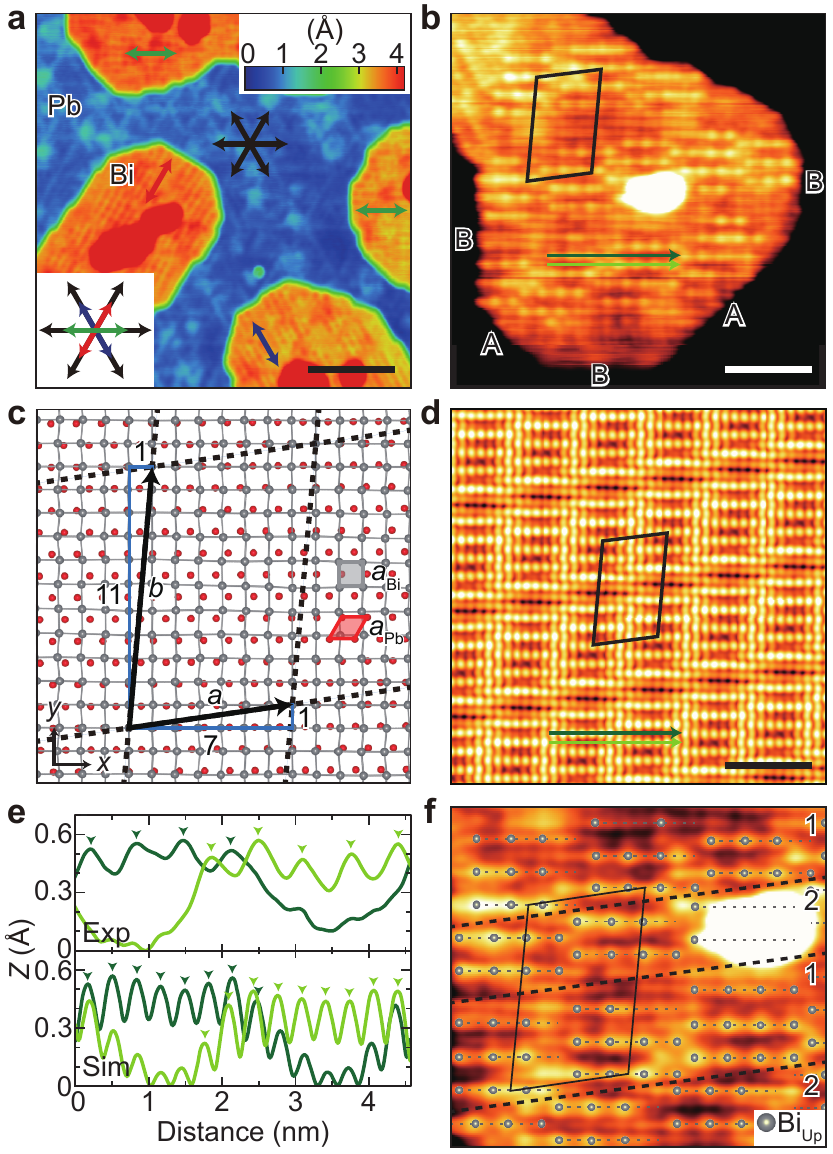}
\caption{\label{fig2} \textbf{Atomic structure of a monolayer Bi/Pb(111) heterostructure. (a) STM image (30 $\times$ 30~ nm$^2$, $V_{\textrm{bias}}$  = $-$100~mV, $I_\textrm{t}$ = 100~pA) of neighboring Bi islands with three different orientations in their atomic rows as indicated by colored arrows. Black arrows indicates the atomic rows of Pb(111). Scale bar, 7~nm. (b) Zoom-in STM image (13 $\times$ 13~nm$^2$, $V_{\textrm{bias}}$  = $-$400~mV, $I_\textrm{t}$ = 100~pA) of a monolayer Bi island. The parallelogram denotes a moir\'e supercell. Scale bar, 3~nm.
(c) Atomic structure of a monolayer Bi(110)/Pb(111) heterostructure optimized in DFT calculations. Grey and red dots are Bi and Pb atoms, respectively. Black arrows denote unit vectors of a moir\'e supercell ($a$ = 22.8~\r{A} and $b$ = 36.3~\r{A}) and the numbers are counts of unit cells ($a_{\textrm{Bi}}$) for guide.  
The unit cells of square-lattice Bi and Pb(111) are indicated ($a_{\textrm{Bi}}$ = 3.3~\r{A}, $a_{\textrm{Pb}}$ = 3.5~\r{A}). The buckling in monolayer Bi is set to zero and the interlayer spacing ($a_{\textrm{Bi-Pb}}$) is 3.2~\r{A}.
(d) DFT simulation of the STM image by integrating LDOS from the Fermi level to $-$0.2~eV for the structure shown in (c). Scale bar, 3~nm.
(e) Line profiles along the two colored arrows in the STM image of (b) and those in the corresponding rows in the simulation of (d). 
(f) Enlarged atomic scale STM image for the same Bi island of (b). Grey dashed lines denotes short stripes. The stronger protrusions are indicated by grey balls (Bi$_{\textrm{up}}$), which are in local $\sqrt{2}\times\sqrt{2}$ units, indicating buckling of Bi atoms, which also exhibit phase shifts as denoted by dashed lines.}}
\end{figure*}

\end{spacing}

%Reason to analyze the buckled bi monolyer/Pb heterostructure
Going into the detailed atomic structure, there exists an important difference between 1 and 2 ML islands, namely, an extra complex periodic modulation is observed in the topography of monolayer islands.
As shown in the STM image and its FFT (upper left quadrant in Figure 1e and Figure 1g), monolayer Bi islands have a local $\sqrt{2}\times\sqrt{2}$ modulation, which is decorated with another modulation of a much larger periodicity.
This modulation appears as short stripes whose contrast is modulated with a periodicity of $a\times b$ $\sim$ 23.0~\r{A} $\times$ 36.0~\r{A} (Figure 2a and parallelogram in Figure 2b).
By resolving the atomic images of both Bi islands and the Pb(111) topmost layer, we confirm that the short stripe along the [100] direction of Bi islands is matched with atomic rows of Pb(111) (inset of Figure 2a). 
This information is put into the DFT calculations to find the interfacial structure. 
Note that we set lattice constants of Bi ($a_{\textrm{Bi}}$ = 3.3~\r{A}) and Pb ($a_{\textrm{Pb}}$ = 3.5~\r{A}) at an interlayer spacing of 3.2~\r{A} without any buckling to reduce the calculational burden.
The optimized structure is shown in Figure 2c and its simulated STM image is shown in Figure 2d. 
One can find a reasonable agreement between the observed and the simulated STM images, which confirms that the superstructure of the monolayer Bi is due to the moir\'e pattern (see black parallelogram in Figure 2b,d). 
Moreover, the 2D FFT analysis for a STM image taken on a larger island shows the superstructure peaks, which are quantitatively consistent with the moir\'e structure (see Figure S4).
The simulation also reveals that the contrast modulation of Bi atoms is mainly due to the strong modulation of local density of states (LDOS) between the Pb top (bright stripes) or hollow sites (dark stripes) of the substrate (see Figure S5).
What is still missing in the present calculation is the local $\sqrt{2}\times\sqrt{2}$ modulation (0.1-0.2~\r{A} in the apparent height) on stripe protrusions in STM images (Figure 2b,e).
The height difference is similar to the buckling of the top layer of the puckered $\alpha$-Bi structure~\cite{Sun2012,Jin2021} (Figure 1a) and the buckling results in the formation of domains with different buckling directions (dashed lines in Figure 2f).
There also exists the atomic protrusions for the position with reduced contrast due to modulated structure, which is more clear in the low-bias image, confirming the $1\times1$ unitcell composed of up and bottom atoms (see domain analysis of Figure 1e in Figure S6 and Figure S7).
As described above, such a mild buckling does not affect significantly the electronic structure of interest here. 
As also shown in Figure 2a, the monolayer Bi islands grow in three different orientations (inset of Figure 2a), due to the triangular symmetry of the substrate Pb(111), and are elongated along the [100] direction.
The elongated islands have both A-type (zigzag and armchair structures) and B-type edges (Figure 2b), in contrast to $\alpha$-Bi films growing in nanoribbon shapes with only A-type edges~\cite{Sun2012,Lu2015,Peng2019}.

\begin{spacing}{1.12} 

\begin{figure*}[htb!]
\centering
\includegraphics[width=164.6mm]{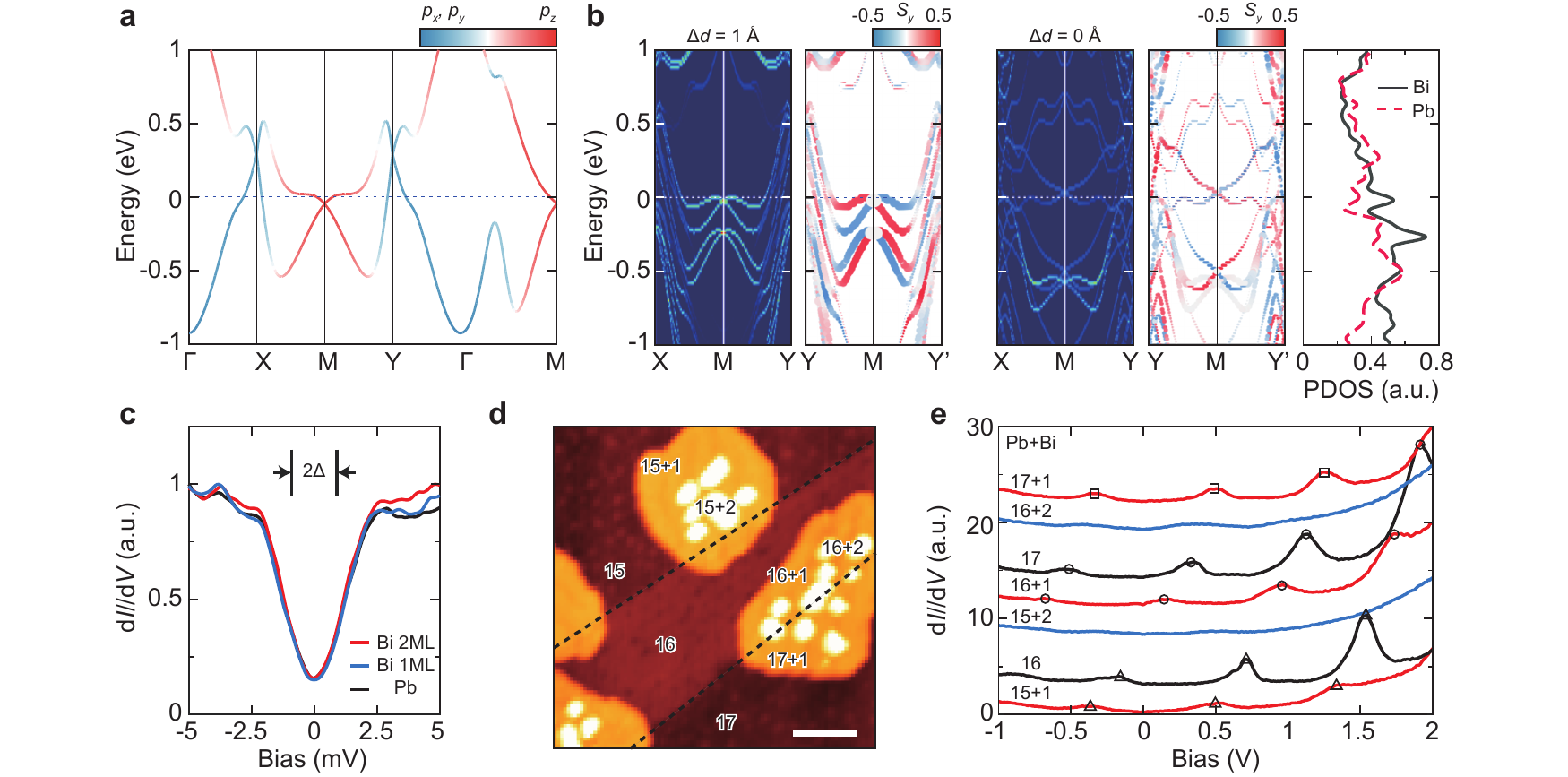}
\caption{\label{fig3} \textbf{Electronic states of a Bi monolayer and a Bi/Pb heterostructure.
(a) Calculated band dispersions of a freestanding square-lattice Bi monolayer.
The $p_{xy}$ and $p_{z}$ orbital contributions are separated with blue and red colors, respectively. 
(b) Calculated band dispersions of the Bi/Pb heterostructure shown in Figure 2c, which are unfolded into the $1\times1$ Brillouin zone. 
The in-plane spin polarization ($S_y$) is indicated by the color scale. The Y' point corresponds to the Y point of the next Brillouin zone.
In order to expose the Bi-Pb interaction, the same calculations are performed for an increased ($\Delta d$ = 1~\r{A}) and the optimized ($\Delta d$ = 0~\r{A}) interlayer spacing. The calculated partial DOS are added for the optimized case.
(c) Tunneling d$I$/d$V$ spectra around the Fermi level showing the superconducting gap 2$\Delta$ ($\Delta$ = 0.9~meV) on a Pb film (10 ML) and Bi monolayer/bilyer islands at 4.3~K with modulation voltage $V_{\textrm{mod}}$ = 100~$\muup$V.
(d) STM image (50 $\times$ 50~nm$^2$, $V_{\textrm{bias}}$ = $-$1~V, $I_\textrm{t}$ = 20~pA) on a Pb(111) film with a thickness of 15, 16, and 17 ML crossing two step edges (dashed lines) of the Si(111) substrate and hosting a few Bi islands. Bi islands have monolayer (+1) and bilayer (+2, brightest parts) heights. The layer thickness is indicated for different parts with consideration of the wetting layer. Scale bar, 10~nm.
(e) Tunneling d$I$/d$V$ spectra ($V_{\textrm{mod}}$ = 10~mV) taken on Pb(111) films of different thickness (black lines), Bi monolayer (red lines), and Bi bilayer islands (blue lines) of (d). Quantum-well states are indicated by symbols. Offsets are given for clarity.
}}
\end{figure*}

\end{spacing}

\subsection{Electronic states of Bi/Pb heterostructure}
In order to investigate the electronic band structure of a monolayer Bi, we perform DFT calculations for freestanding square-lattice monolayer of Bi. 
Calculated band dispersions (Figure 3a) of a freestanding square-lattice monolayer (4.7~\r{A} $\times$ 4.7~\r{A} and buckling height $h$ = 0.73~\r{A}) are symmetric along $x$ and $y$ momenta with respect to the M point of its Brillouin zone due to C$_4$ symmetry and feature three Dirac crossings at M, X, and Y points.
While the buckling height is sensitive to in-plane lattice constant~\cite{Jin2021}, these Dirac crossings are preserved as protected by the symmetry (see Figure S8).
In contrast to $\beta$- and $\alpha$-Bi monolayers, this is a metallic system. 
This difference is due to the presence of $p_z$ dangling bonds, which are responsible for the Dirac bands.
While the Dirac points at X and Y are located in at 0.3~eV above Fermi level to become inaccessible in photoemission and transport measurements, the Dirac point at M is located just below the Fermi level and is protected by two mirror planes with a four-fold degeneracy (no spin polarization)~\cite{Bian2014,Kowalczyk2017}.
This band structure is overall very similar to the case of the compressed monolayer Bi~\cite{Kowalczyk2017}.
We further calculate band structures of a Bi/Pb supercell including the substrate.
The effect of the substrate can be traced by changing the Bi-Pb interlayer separation (see Figure 3b and Figure S9).
The main change is that the degenerated Dirac crossings at M points split with an energy separation of eventually 0.5~eV at the equilibrium distance ($\Delta d$ = 0~\r{A}).
The upper Dirac crossing is located very close to the Fermi level, becoming similar to that of the freestanding film, but the other Dirac cone is located well within valence bands.
Very importantly, a substantial Rashba-type spin-splitting occurs as shown in the Figure 3b due to the strong SOC of Bi and Pb and to the breaking of sublattice symmetry by the Pb substrate (Figure 2c).
The strong interfacial interaction also induces a substantial Pb contribution to the Rashba bands (see Figures S9 and S10 for the evolution of the Rashba-split bands).
These behaviors are largely different from the Rashba behaviour of weakly-interacting graphene heterointerfaces~\cite{Marchenko2012}.

The calculated DOS (Figure 3b) exhibits metallic property with contributions of both Bi and Pb at Fermi level.  
The metallic property of the Bi monolayer is confirmed by the differential conductance (d$I$/d$V$) measurements and corroborated by the proximity-induced superconductivity. 
On Pb(111) films ($T_c$ = 7.2~K in Bulk), the superconducting gap is clearly observed at 4.3 K as previously reported ~\cite{Liu2011_1}. By comparing the zero bias conductance in thin films~\cite{Eom2006}, almost the same size of a gap ($\Delta$ = 0.9~meV) is estimated in both 1 and 2 ML Bi islands as shown in Figure 3c [$\Delta$~($T$ = 0~K) = 1.2~meV].
Note also that a similar superconducting gap of $\sim$1.0~meV was observed for puckered Bi(110) films on NbSe$_2$~\cite{Peng2019}.

In addition to the superconducting gap of a very small energy scale, we observe large energy-scale spectral features on Bi/Pb thin films (Figure 3d) as shown in Figure 3e. Basically, these equally-spaced spectral features originate from the well known quantum well states of Pb(111) films (black spectra) and their quantization energies agree well with the previous photoemission spectroscopy and scanning tunneling spectroscopy observations~\cite{Mans2002,Wei2002,Hong2009,Pan2011}. 
With one monolayer of Bi on top of Pb(111) films, the quantum well state energies are observed to shift. 
This behavior was observed previously, which is due to the change of the quantization condition by the presence of one more metallic layer on top~\cite{Vyalikh2003,Pan2020}. 
On the other hand, the peak intensities of Pb quantum well states are reduced for a monolayer Bi and further for a second layer Bi as naturally expected~\cite{Chiang2000}.

\subsection{Possibility of topological superconductivity (TSC)}
The Rashba-split bands are degenerates at the M point, forming the Kramers pairs, whose essential physics can be described by an effective tight-binding model using Bi $p_{z}$ orbitals (first panel of Figure 4a). 
This band structure can lead to TSC when combined with exchange field and superconducting pairing interaction (see Supporting Information Part \Romannum{2} for the model Hamiltonian). 
For the case of a simple s-wave superconductor with the pairing gap of $\Delta_{\textrm{s}}$ = 1~meV (similar to the experimental value) and without an exchange field, we can obtain the topologically trivial dispersion of superconducting quasiparticles (second panel of Figure 4a). 
When the out-of-plane exchange field $B_{z}$ is turned on, spin-up and spin-down bands are split and mixed. With increasing $B_{z}$, the gap at the M point decreases to close at a critical value of $B_{z}$ = 1~meV and then reopens, indicating a topological phase transition (third and fourth panels of Figure 4a). 
The Chern number after the transition is calculated from the Berry curvature of the bands to be 1 (see Supporting Information Part \Romannum{2}), which belongs to the topological class D~\cite{Schnyder2008}.
Further, we investigate the topological phase diagram by tracing the band gap at the M point. 
The phase boundary between the normal SC and TSC is determined as indicated the dashed line in Figure 4b, i.e., $B_{z}$ $\geq$ $\Delta_{\textrm{s}}$~\cite{Jin2019_1,Zhang2021}. 
The nontrivial order can also be manifested by the presence of Majorana edge states within the superconducting gap as presented by the zero energy mode in the energy spectra (Figure 4c).
Therefore, the present Bi/Pb(111) system would be a promising candidate to observe a Majorana zero mode, for example, around magnetic impurities and adsorbates or on island step edge under an uniform magnetic field.
For the latter case, the critical field strength to induce a topological phase can be estimated in terms of the Zeeman gap, $B_{z}$ = $g\mu_{\textrm{B}}B$/2, where $g$ is the Land\'e $g$-factor, $\mu_{\textrm{B}}$ is the Bohr magneton, and $B$ is the external magnetic field. Using the value of the Land\'e $g$-factor for Bi of 63.2~\cite{Verdun1976}, the critical magnetic field strength for a Zeeman gap of 1~meV is approximately 0.55~T, which is smaller than the upper critical field range of 1-4~T of Pb films~\cite{Hsu1993,Liu2018}. This indicates that the necessary magnetic field conditions for realizing Majorana modes are achievable.

\begin{spacing}{1.12} 

\begin{figure*}[t!]
\centering
\includegraphics[width=84.6mm]{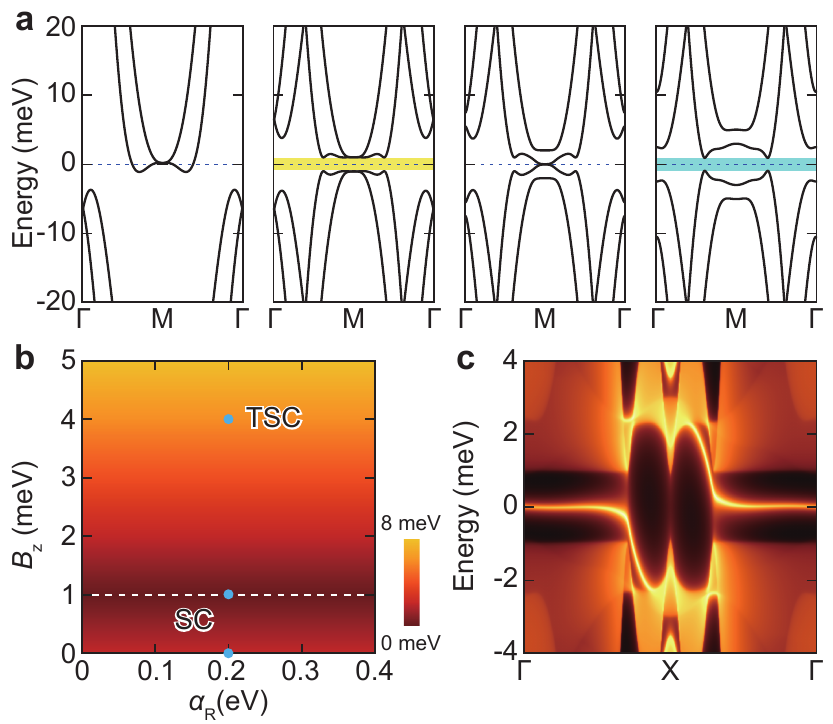}
\caption{\label{fig4} \textbf{Tight-binding (TB) calculation for Bi Rashba bands with the superconducting gap. (a) First panel: The calculated band dispersions of the TB Hamiltonian with hopping parameters $t_{1}$= $-$0.103~eV, $t_{2}$ = $-$0.074~eV, $t_{3}$ = 0.057~eV, chemical potential $\mu$ = 0.13~eV, and Rashba coupling $\alpha_\textrm{R}$ = 0.2~eV (see Supporting Information Part \Romannum{2}). Second to forth panel: The calculated band dispersions after including the pairing interaction with the superconducting gap of $\Delta_{\textrm{s}}$ = 1~meV under the magnetic field $B_{z}$ of 0 (second panel), 1 (third panel), and 4~meV (fourth panel), respectively. The colored stripe indicates the superconducting gap. (b) Topological phase diagram in the parameter space of $\alpha_\textrm{R}$ and $B_{z}$. The color scale indicates the size of the band gap at M point. The band structure parameters for (a) are marked by the dots. The dashed line is the boundary between normal SC and TSC phases. (c) The calculated edge state spectral function of a semi-infinite Bi monolayer in the TSC phase.}}
\end{figure*}

\end{spacing}

%CONCLUSION
%%%%%%%%%%%%%%%%%%%%%%%%%%%%%%%%%%%%%%%%%%%%%%%%%%%%%%%%%%%%%%%%%%%%%%%%%%%%%%%%%%%%%%%%
%%%%%%%%%%%%%%%%%%%%%%%%%%%%%%%%%%%%%%%%%%%%%%%%%%%%%%%%%%%%%%%%%%%%%%%%%%%%%%%%%%%%%%%%
%%%%%%%%%%%%%%%%%%%%%%%%%%%%%%%%%%%%%%%%%%%%%%%%%%%%%%%%%%%%%%%%%%%%%%%%%%%%%%%%%%%%%%%%

\section{CONCLUSIONS}
\noindent
In summary, we fabricate a nearly flat Bi(110) monolayer and bilayer with a distinct square lattice on Pb(111) thin films. 
Our STM study reveals that the monolayer has two mirror planes and C$_4$ symmetry, whose unit length is comparable with that of Bi(110) films. 
The monolayer film exhibits a stripy moir\'e superstructure, whose atomic structure is determined by STM and DFT calculations.
DFT calculations predict that the square-lattice Bi monolayer features spin-split Dirac crossing due to the strong SOC at the Fermi level and the sublattice symmetry breaking by the strong interfacial interaction.
The Bi monolayer and bilayer are shown to be superconducting at 4.3~K through the proximity coupling with the superconducting Pb substrate. 
The strongly spin-split Dirac band of Bi monolayer is expected to induce a topological superconducting state with the exchange field of magnetic dopants or a uniform magnetic field in our tight-binding model calculations. 
An intriguing material platform with 2D Dirac bands, strong SOC, superconductivity, and the moir\'e superstructure is, thus, established in this work. 

%METHODS
%%%%%%%%%%%%%%%%%%%%%%%%%%%%%%%%%%%%%%%%%%%%%%%%%%%%%%%%%%%%%%%%%%%%%%%%%%%%%%%%%%%%%%%%
%%%%%%%%%%%%%%%%%%%%%%%%%%%%%%%%%%%%%%%%%%%%%%%%%%%%%%%%%%%%%%%%%%%%%%%%%%%%%%%%%%%%%%%%
%%%%%%%%%%%%%%%%%%%%%%%%%%%%%%%%%%%%%%%%%%%%%%%%%%%%%%%%%%%%%%%%%%%%%%%%%%%%%%%%%%%%%%%%

\section{METHODS}
\textbf{Molecular beam epitaxy (MBE) Growth.}
Pb films of about seven monolayers nominally were deposited onto a Si(111)-7$\times$7 surface held at room temperature, which were subsequently cooled down to 86~K for the deposition (12 min) of Bi from a standard Knudsen cell at 703~K. 

\textbf{STM and STS measurements.}
Experiments were performed in ultra high vacuum cryogenic STM systems (SPECS and Unisoku) at 78 and 4.3~K in the constant-current mode using mechanically sharpen Pt-Ir tips.
A lock-in technique was used to measure d$I$/d$V$ spectra with a modulation of 1~kHz

\textbf{Theoretical calculations.}
DFT calculations were carried out using the Vienna \textit{ab initio} simulation package within the generalized-gradient approximation (GGA)~\cite{Kresse1996,Perdew1996} with the projector augmented wave method. 
Calculations are carried out up to the kinetic energy cutoff of 400~eV on a $21\times21\times1$ and $3\times3\times1$ k-point mesh for freestanding monolayer Bi and Bi/Pb heterostructures, respectively. 
All structures are fully optimized until residual forces are less than 0.02~eV/~\r{A} with SOC included in a self-consistent way.

%ASSOCIATED CONTENT
%%%%%%%%%%%%%%%%%%%%%%%%%%%%%%%%%%%%%%%%%%%%%%%%%%%%%%%%%%%%%%%%%%%%%%%%%%%%%%%%%%%%%%%%
%%%%%%%%%%%%%%%%%%%%%%%%%%%%%%%%%%%%%%%%%%%%%%%%%%%%%%%%%%%%%%%%%%%%%%%%%%%%%%%%%%%%%%%%
%%%%%%%%%%%%%%%%%%%%%%%%%%%%%%%%%%%%%%%%%%%%%%%%%%%%%%%%%%%%%%%%%%%%%%%%%%%%%%%%%%%%%%%%

\section{ASSOCIATED CONTENT}
\subsection{Supporting Information}
\noindent
The Supporting Information is available free of charge at https://%pubs.acs.org.

Results of room-temperature deposition of bismuth, tight-binding calculation of an effective topological superconductor model, FFT analysis of moir\'e pattern, electronic structure depending on Bi position, Domain analysis on 1ML, low-bias topographic image, band dispersions of Bi monolayer depending on buckling heights, band dispersions of Bi/Pb at different interfacial distance. and distribution of total Berry curvature for the states below SC gap (PDF)

%AUTHOR INFORMATION
%%%%%%%%%%%%%%%%%%%%%%%%%%%%%%%%%%%%%%%%%%%%%%%%%%%%%%%%%%%%%%%%%%%%%%%%%%%%%%%%%%%%%%%%
%%%%%%%%%%%%%%%%%%%%%%%%%%%%%%%%%%%%%%%%%%%%%%%%%%%%%%%%%%%%%%%%%%%%%%%%%%%%%%%%%%%%%%%%
%%%%%%%%%%%%%%%%%%%%%%%%%%%%%%%%%%%%%%%%%%%%%%%%%%%%%%%%%%%%%%%%%%%%%%%%%%%%%%%%%%%%%%%%

\section{AUTHOR INFORMATION}
\subsection{Corresponding Author}
\textbf{Han Woong Yeom}, $^*$Email: yeom@postech.ac.kr

\subsection{Author contributions}
\noindent
H.W.Y. conceived the research idea and plan. E.O. performed the STM experiments. K.-H.J. performed the DFT calculations and the model calculation. H.W.Y. and E.O. prepared the manuscript. All authors discussed the results and commented on the paper.

\subsection{Author contributions}
\noindent
$^{\parallel}$E.O. and K.-H.J. contributed equally to this work.

\subsection{Notes}
\noindent
The authors declare no competing financial interest.

%ACKNOWLEDGNMENTS
%%%%%%%%%%%%%%%%%%%%%%%%%%%%%%%%%%%%%%%%%%%%%%%%%%%%%%%%%%%%%%%%%%%%%%%%%%%%%%%%%%%%%%%%
%%%%%%%%%%%%%%%%%%%%%%%%%%%%%%%%%%%%%%%%%%%%%%%%%%%%%%%%%%%%%%%%%%%%%%%%%%%%%%%%%%%%%%%%
%%%%%%%%%%%%%%%%%%%%%%%%%%%%%%%%%%%%%%%%%%%%%%%%%%%%%%%%%%%%%%%%%%%%%%%%%%%%%%%%%%%%%%%%

\section{ACKNOWLEDGNMENTS}
\noindent
This work was supported by the Institute for Basic Science (Grant No. IBS-R014-D1). K.-H.J. is supported by the Institute for Basic Science (Grant No. IBS-R014-Y1).

%Reference
%%%%%%%%%%%%%%%%%%%%%%%%%%%%%%%%%%%%%%%%%%%%%%%%%%%%%%%%%%%%%%%%%%%%%%%%%%%%%%%%%%%%%%%%
%%%%%%%%%%%%%%%%%%%%%%%%%%%%%%%%%%%%%%%%%%%%%%%%%%%%%%%%%%%%%%%%%%%%%%%%%%%%%%%%%%%%%%%%
%%%%%%%%%%%%%%%%%%%%%%%%%%%%%%%%%%%%%%%%%%%%%%%%%%%%%%%%%%%%%%%%%%%%%%%%%%%%%%%%%%%%%%%%

\bibliography{Main}

\providecommand{\latin}[1]{#1}
\makeatletter
\providecommand{\doi}
  {\begingroup\let\do\@makeother\dospecials
  \catcode`\{=1 \catcode`\}=2 \doi@aux}
\providecommand{\doi@aux}[1]{\endgroup\texttt{#1}}
\makeatother
\providecommand*\mcitethebibliography{\thebibliography}
\csname @ifundefined\endcsname{endmcitethebibliography}
  {\let\endmcitethebibliography\endthebibliography}{}
\begin{mcitethebibliography}{67}
\providecommand*\natexlab[1]{#1}
\providecommand*\mciteSetBstSublistMode[1]{}
\providecommand*\mciteSetBstMaxWidthForm[2]{}
\providecommand*\mciteBstWouldAddEndPuncttrue
  {\def\EndOfBibitem{\unskip.}}
\providecommand*\mciteBstWouldAddEndPunctfalse
  {\let\EndOfBibitem\relax}
\providecommand*\mciteSetBstMidEndSepPunct[3]{}
\providecommand*\mciteSetBstSublistLabelBeginEnd[3]{}
\providecommand*\EndOfBibitem{}
\mciteSetBstSublistMode{f}
\mciteSetBstMaxWidthForm{subitem}{(\alph{mcitesubitemcount})}
\mciteSetBstSublistLabelBeginEnd
  {\mcitemaxwidthsubitemform\space}
  {\relax}
  {\relax}

\bibitem[Bhimanapati \latin{et~al.}(2015)Bhimanapati, Lin, Meunier, Jung, Cha,
  Das, Xiao, Son, Strano, Cooper, Liang, Louie, Ringe, Zhou, Kim, Naik,
  Sumpter, Terrones, Xia, Wang, Zhu, Akinwande, Alem, Schuller, Schaak,
  Terrones, and Robinson]{Bhimanapati2015}
Bhimanapati,~G.~R. \latin{et~al.}  Recent advances in two-dimensional materials
  beyond graphene. \emph{ACS Nano} \textbf{2015}, \emph{9}, 11509--11539\relax
\mciteBstWouldAddEndPuncttrue
\mciteSetBstMidEndSepPunct{\mcitedefaultmidpunct}
{\mcitedefaultendpunct}{\mcitedefaultseppunct}\relax
\EndOfBibitem
\bibitem[Liu and Hersam(2019)Liu, and Hersam]{Liu2019}
Liu,~X.; Hersam,~M.~C. 2D materials for quantum information science.
  \emph{Nature Reviews Materials} \textbf{2019}, \emph{4}, 669--684\relax
\mciteBstWouldAddEndPuncttrue
\mciteSetBstMidEndSepPunct{\mcitedefaultmidpunct}
{\mcitedefaultendpunct}{\mcitedefaultseppunct}\relax
\EndOfBibitem
\bibitem[Kou \latin{et~al.}(2017)Kou, Ma, Sun, Heine, and Chen]{Kou2017}
Kou,~L.; Ma,~Y.; Sun,~Z.; Heine,~T.; Chen,~C. Two-dimensional topological
  insulators: Progress and prospects. \emph{The Journal of Physical Chemistry
  Letters} \textbf{2017}, \emph{8}, 1905--1919\relax
\mciteBstWouldAddEndPuncttrue
\mciteSetBstMidEndSepPunct{\mcitedefaultmidpunct}
{\mcitedefaultendpunct}{\mcitedefaultseppunct}\relax
\EndOfBibitem
\bibitem[Ahn(2020)]{Ahn2020}
Ahn,~E.~C. 2D materials for spintronic devices. \emph{npj 2D Materials and
  Applications} \textbf{2020}, \emph{4}, 17\relax
\mciteBstWouldAddEndPuncttrue
\mciteSetBstMidEndSepPunct{\mcitedefaultmidpunct}
{\mcitedefaultendpunct}{\mcitedefaultseppunct}\relax
\EndOfBibitem
\bibitem[J{\"a}ck \latin{et~al.}(2021)J{\"a}ck, Xie, and Yazdani]{Jack2021}
J{\"a}ck,~B.; Xie,~Y.; Yazdani,~A. Detecting and distinguishing majorana zero
  modes with the scanning tunnelling microscope. \emph{Nature Reviews Physics}
  \textbf{2021}, \emph{3}, 541--554\relax
\mciteBstWouldAddEndPuncttrue
\mciteSetBstMidEndSepPunct{\mcitedefaultmidpunct}
{\mcitedefaultendpunct}{\mcitedefaultseppunct}\relax
\EndOfBibitem
\bibitem[Sierra \latin{et~al.}(2021)Sierra, Fabian, Kawakami, Roche, and
  Valenzuela]{Sierra2021}
Sierra,~J.~F.; Fabian,~J.; Kawakami,~R.~K.; Roche,~S.; Valenzuela,~S.~O. Van
  der waals heterostructures for spintronics and opto-spintronics. \emph{Nature
  Nanotechnology} \textbf{2021}, \emph{16}, 856--868\relax
\mciteBstWouldAddEndPuncttrue
\mciteSetBstMidEndSepPunct{\mcitedefaultmidpunct}
{\mcitedefaultendpunct}{\mcitedefaultseppunct}\relax
\EndOfBibitem
\bibitem[Zhang \latin{et~al.}(2016)Zhang, Xie, Li, Yan, Li, Kan, Liu, Chen, and
  Zeng]{Zhang2016}
Zhang,~S.; Xie,~M.; Li,~F.; Yan,~Z.; Li,~Y.; Kan,~E.; Liu,~W.; Chen,~Z.;
  Zeng,~H. Semiconducting group 15 monolayers: A broad range of band gaps and
  high carrier mobilities. \emph{Angewandte Chemie International Edition}
  \textbf{2016}, \emph{55}, 1666--1669\relax
\mciteBstWouldAddEndPuncttrue
\mciteSetBstMidEndSepPunct{\mcitedefaultmidpunct}
{\mcitedefaultendpunct}{\mcitedefaultseppunct}\relax
\EndOfBibitem
\bibitem[Zhang \latin{et~al.}(2018)Zhang, Guo, Chen, Wang, Gao,
  G{\'o}mez-Herrero, Ares, Zamora, Zhu, and Zeng]{Zhang2018}
Zhang,~S.; Guo,~S.; Chen,~Z.; Wang,~Y.; Gao,~H.; G{\'o}mez-Herrero,~J.;
  Ares,~P.; Zamora,~F.; Zhu,~Z.; Zeng,~H. Recent progress in 2D group-VA
  semiconductors: from theory to experiment. \emph{Chemical Society Reviews}
  \textbf{2018}, \emph{47}, 982--1021\relax
\mciteBstWouldAddEndPuncttrue
\mciteSetBstMidEndSepPunct{\mcitedefaultmidpunct}
{\mcitedefaultendpunct}{\mcitedefaultseppunct}\relax
\EndOfBibitem
\bibitem[Wang \latin{et~al.}(2015)Wang, Pandey, and Karna]{Wang2015}
Wang,~G.; Pandey,~R.; Karna,~S.~P. Atomically thin group V elemental films:
  Theoretical investigations of antimonene allotropes. \emph{ACS Applied
  Materials {$\&$} Interfaces} \textbf{2015}, \emph{7}, 11490--11496\relax
\mciteBstWouldAddEndPuncttrue
\mciteSetBstMidEndSepPunct{\mcitedefaultmidpunct}
{\mcitedefaultendpunct}{\mcitedefaultseppunct}\relax
\EndOfBibitem
\bibitem[M{\"a}rkl \latin{et~al.}(2017)M{\"a}rkl, Kowalczyk, Ster, Mahajan,
  Pirie, Ahmed, Bian, Wang, Chiang, and Brown]{Markl2018}
M{\"a}rkl,~T.; Kowalczyk,~P.~J.; Ster,~M.~L.; Mahajan,~I.~V.; Pirie,~H.;
  Ahmed,~Z.; Bian,~G.; Wang,~X.; Chiang,~T.-C.; Brown,~S.~A. Engineering
  multiple topological phases in nanoscale Van der Waals heterostructures:
  realisation of $\alpha$-antimonene. \emph{2D Materials} \textbf{2017},
  \emph{5}, 011002\relax
\mciteBstWouldAddEndPuncttrue
\mciteSetBstMidEndSepPunct{\mcitedefaultmidpunct}
{\mcitedefaultendpunct}{\mcitedefaultseppunct}\relax
\EndOfBibitem
\bibitem[Hirahara \latin{et~al.}(2011)Hirahara, Bihlmayer, Sakamoto, Yamada,
  Miyazaki, Kimura, Bl\"ugel, and Hasegawa]{Hirahara2011}
Hirahara,~T.; Bihlmayer,~G.; Sakamoto,~Y.; Yamada,~M.; Miyazaki,~H.;
  Kimura,~S.-i.; Bl\"ugel,~S.; Hasegawa,~S. Interfacing 2D and 3D topological
  insulators: Bi(111) bilayer on Bi$_{2}$Te$_{3}$. \emph{Phys. Rev. Lett.}
  \textbf{2011}, \emph{107}, 166801\relax
\mciteBstWouldAddEndPuncttrue
\mciteSetBstMidEndSepPunct{\mcitedefaultmidpunct}
{\mcitedefaultendpunct}{\mcitedefaultseppunct}\relax
\EndOfBibitem
\bibitem[Li \latin{et~al.}(2022)Li, Yu, Zhang, Li, Qiao, Peng, Dong, Wang, Ma,
  Xiao, and Yao]{Li2022}
Li,~J.; Yu,~K.; Zhang,~X.; Li,~Y.; Qiao,~L.; Peng,~X.; Dong,~X.; Wang,~Z.;
  Ma,~J.; Xiao,~W.; Yao,~Y. Controllable growth of $\alpha$- and
  $\beta$-antimonene by interfacial strain. \emph{The Journal of Physical
  Chemistry C} \textbf{2022}, \emph{126}, 5022--5027\relax
\mciteBstWouldAddEndPuncttrue
\mciteSetBstMidEndSepPunct{\mcitedefaultmidpunct}
{\mcitedefaultendpunct}{\mcitedefaultseppunct}\relax
\EndOfBibitem
\bibitem[Bian \latin{et~al.}(2009)Bian, Miller, and Chiang]{Bian2009}
Bian,~G.; Miller,~T.; Chiang,~T.-C. Electronic structure and surface-mediated
  metastability of Bi films on Si(111)-7$\times$7 studied by angle-resolved
  photoemission spectroscopy. \emph{Phys. Rev. B} \textbf{2009}, \emph{80},
  245407\relax
\mciteBstWouldAddEndPuncttrue
\mciteSetBstMidEndSepPunct{\mcitedefaultmidpunct}
{\mcitedefaultendpunct}{\mcitedefaultseppunct}\relax
\EndOfBibitem
\bibitem[Scott \latin{et~al.}(2005)Scott, Kral, and Brown]{Scott2005}
Scott,~S.; Kral,~M.; Brown,~S. A crystallographic orientation transition and
  early stage growth characteristics of thin Bi films on HOPG. \emph{Surface
  Science} \textbf{2005}, \emph{587}, 175--184\relax
\mciteBstWouldAddEndPuncttrue
\mciteSetBstMidEndSepPunct{\mcitedefaultmidpunct}
{\mcitedefaultendpunct}{\mcitedefaultseppunct}\relax
\EndOfBibitem
\bibitem[Nagao \latin{et~al.}(2004)Nagao, Sadowski, Saito, Yaginuma, Fujikawa,
  Kogure, Ohno, Hasegawa, Hasegawa, and Sakurai]{Nagao2004}
Nagao,~T.; Sadowski,~J.~T.; Saito,~M.; Yaginuma,~S.; Fujikawa,~Y.; Kogure,~T.;
  Ohno,~T.; Hasegawa,~Y.; Hasegawa,~S.; Sakurai,~T. Nanofilm allotrope and
  phase transformation of ultrathin Bi film on Si(111)-7$\times$7. \emph{Phys.
  Rev. Lett.} \textbf{2004}, \emph{93}, 105501\relax
\mciteBstWouldAddEndPuncttrue
\mciteSetBstMidEndSepPunct{\mcitedefaultmidpunct}
{\mcitedefaultendpunct}{\mcitedefaultseppunct}\relax
\EndOfBibitem
\bibitem[Hofmann(2006)]{hoffman2006}
Hofmann,~P. The surfaces of bismuth: Structural and electronic properties.
  \emph{Progress in Surface Science} \textbf{2006}, \emph{81}, 191--245\relax
\mciteBstWouldAddEndPuncttrue
\mciteSetBstMidEndSepPunct{\mcitedefaultmidpunct}
{\mcitedefaultendpunct}{\mcitedefaultseppunct}\relax
\EndOfBibitem
\bibitem[Wada \latin{et~al.}(2011)Wada, Murakami, Freimuth, and
  Bihlmayer]{Wada2011}
Wada,~M.; Murakami,~S.; Freimuth,~F.; Bihlmayer,~G. Localized edge states in
  two-dimensional topological insulators: Ultrathin Bi films. \emph{Phys. Rev.
  B} \textbf{2011}, \emph{83}, 121310\relax
\mciteBstWouldAddEndPuncttrue
\mciteSetBstMidEndSepPunct{\mcitedefaultmidpunct}
{\mcitedefaultendpunct}{\mcitedefaultseppunct}\relax
\EndOfBibitem
\bibitem[Yeom \latin{et~al.}(2016)Yeom, Jin, and Jhi]{Yeom2016}
Yeom,~H.~W.; Jin,~K.-H.; Jhi,~S.-H. Topological fate of edge states of single
  Bi bilayer on Bi(111). \emph{Phys. Rev. B} \textbf{2016}, \emph{93},
  075435\relax
\mciteBstWouldAddEndPuncttrue
\mciteSetBstMidEndSepPunct{\mcitedefaultmidpunct}
{\mcitedefaultendpunct}{\mcitedefaultseppunct}\relax
\EndOfBibitem
\bibitem[Murakami(2006)]{Murakami2006}
Murakami,~S. Quantum spin hall effect and enhanced magnetic response by
  spin-orbit coupling. \emph{Phys. Rev. Lett.} \textbf{2006}, \emph{97},
  236805\relax
\mciteBstWouldAddEndPuncttrue
\mciteSetBstMidEndSepPunct{\mcitedefaultmidpunct}
{\mcitedefaultendpunct}{\mcitedefaultseppunct}\relax
\EndOfBibitem
\bibitem[Liu \latin{et~al.}(2011)Liu, Liu, Wu, Duan, Liu, and Wu]{Liu2011}
Liu,~Z.; Liu,~C.-X.; Wu,~Y.-S.; Duan,~W.-H.; Liu,~F.; Wu,~J. Stable nontrivial
  ${Z}_{2}$ topology in ultrathin Bi(111) films: A first-principles study.
  \emph{Phys. Rev. Lett.} \textbf{2011}, \emph{107}, 136805\relax
\mciteBstWouldAddEndPuncttrue
\mciteSetBstMidEndSepPunct{\mcitedefaultmidpunct}
{\mcitedefaultendpunct}{\mcitedefaultseppunct}\relax
\EndOfBibitem
\bibitem[Drozdov \latin{et~al.}(2014)Drozdov, Alexandradinata, Jeon,
  Nadj-Perge, Ji, Cava, Andrei~Bernevig, and Yazdani]{Drozdov2014}
Drozdov,~I.~K.; Alexandradinata,~A.; Jeon,~S.; Nadj-Perge,~S.; Ji,~H.;
  Cava,~R.~J.; Andrei~Bernevig,~B.; Yazdani,~A. One-dimensional topological
  edge states of bismuth bilayers. \emph{Nature Physics} \textbf{2014},
  \emph{10}, 664--669\relax
\mciteBstWouldAddEndPuncttrue
\mciteSetBstMidEndSepPunct{\mcitedefaultmidpunct}
{\mcitedefaultendpunct}{\mcitedefaultseppunct}\relax
\EndOfBibitem
\bibitem[Yang \latin{et~al.}(2012)Yang, Miao, Wang, Yao, Zhu, Song, Wang, Xu,
  Fedorov, Sun, Zhang, Liu, Liu, Qian, Gao, and Jia]{Yang2012}
Yang,~F. \latin{et~al.}  Spatial and energy distribution of topological edge
  states in single Bi(111) bilayer. \emph{Phys. Rev. Lett.} \textbf{2012},
  \emph{109}, 016801\relax
\mciteBstWouldAddEndPuncttrue
\mciteSetBstMidEndSepPunct{\mcitedefaultmidpunct}
{\mcitedefaultendpunct}{\mcitedefaultseppunct}\relax
\EndOfBibitem
\bibitem[Chen \latin{et~al.}(2012)Chen, Peng, Zhang, Wang, He, Ma, and
  Xue]{Chen2012}
Chen,~M.; Peng,~J.-P.; Zhang,~H.-M.; Wang,~L.-L.; He,~K.; Ma,~X.-C.; Xue,~Q.-K.
  Molecular beam epitaxy of bilayer Bi(111) films on topological insulator
  Bi$_{2}$Te$_{3}$: A scanning tunneling microscopy study. \emph{Applied
  Physics Letters} \textbf{2012}, \emph{101}, 081603\relax
\mciteBstWouldAddEndPuncttrue
\mciteSetBstMidEndSepPunct{\mcitedefaultmidpunct}
{\mcitedefaultendpunct}{\mcitedefaultseppunct}\relax
\EndOfBibitem
\bibitem[Yao \latin{et~al.}(2016)Yao, Zhu, Han, Guan, Liu, Qian, and
  Jia]{Yao2016}
Yao,~M.-Y.; Zhu,~F.; Han,~C.~Q.; Guan,~D.~D.; Liu,~C.; Qian,~D.; Jia,~J.-f.
  Topologically nontrivial bismuth(111) thin films. \emph{Scientific Reports}
  \textbf{2016}, \emph{6}, 21326\relax
\mciteBstWouldAddEndPuncttrue
\mciteSetBstMidEndSepPunct{\mcitedefaultmidpunct}
{\mcitedefaultendpunct}{\mcitedefaultseppunct}\relax
\EndOfBibitem
\bibitem[Sun \latin{et~al.}(2017)Sun, Wang, Zhu, Wang, Ma, Xu, Liao, Lu, Gao,
  Li, Liu, Qian, Guan, and Jia]{Sun2017}
Sun,~H.-H.; Wang,~M.-X.; Zhu,~F.; Wang,~G.-Y.; Ma,~H.-Y.; Xu,~Z.-A.; Liao,~Q.;
  Lu,~Y.; Gao,~C.-L.; Li,~Y.-Y.; Liu,~C.; Qian,~D.; Guan,~D.; Jia,~J.-F.
  Coexistence of topological edge state and superconductivity in bismuth
  ultrathin film. \emph{Nano Letters} \textbf{2017}, \emph{17},
  3035--3039\relax
\mciteBstWouldAddEndPuncttrue
\mciteSetBstMidEndSepPunct{\mcitedefaultmidpunct}
{\mcitedefaultendpunct}{\mcitedefaultseppunct}\relax
\EndOfBibitem
\bibitem[Kim \latin{et~al.}(2014)Kim, Jin, Park, Kim, Jhi, Kim, and
  Yeom]{Kim2014}
Kim,~S.~H.; Jin,~K.-H.; Park,~J.; Kim,~J.~S.; Jhi,~S.-H.; Kim,~T.-H.;
  Yeom,~H.~W. Edge and interfacial states in a two-dimensional topological
  insulator: Bi(111) bilayer on Bi$_{2}$Te$_{2}$Se. \emph{Phys. Rev. B}
  \textbf{2014}, \emph{89}, 155436\relax
\mciteBstWouldAddEndPuncttrue
\mciteSetBstMidEndSepPunct{\mcitedefaultmidpunct}
{\mcitedefaultendpunct}{\mcitedefaultseppunct}\relax
\EndOfBibitem
\bibitem[Peng \latin{et~al.}(2018)Peng, Xian, Tang, Rubio, Zhang, Zhang, and
  Fu]{Peng2018}
Peng,~L.; Xian,~J.-J.; Tang,~P.; Rubio,~A.; Zhang,~S.-C.; Zhang,~W.; Fu,~Y.-S.
  Visualizing topological edge states of single and double bilayer Bi supported
  on multibilayer Bi(111) films. \emph{Phys. Rev. B} \textbf{2018}, \emph{98},
  245108\relax
\mciteBstWouldAddEndPuncttrue
\mciteSetBstMidEndSepPunct{\mcitedefaultmidpunct}
{\mcitedefaultendpunct}{\mcitedefaultseppunct}\relax
\EndOfBibitem
\bibitem[Reis \latin{et~al.}(2017)Reis, Li, Dudy, Bauernfeind, Glass, Hanke,
  Thomale, Schäfer, and Claessen]{Reis2017}
Reis,~F.; Li,~G.; Dudy,~L.; Bauernfeind,~M.; Glass,~S.; Hanke,~W.; Thomale,~R.;
  Schäfer,~J.; Claessen,~R. Bismuthene on a SiC substrate: A candidate for a
  high-temperature quantum spin hall material. \emph{Science} \textbf{2017},
  \emph{357}, 287--290\relax
\mciteBstWouldAddEndPuncttrue
\mciteSetBstMidEndSepPunct{\mcitedefaultmidpunct}
{\mcitedefaultendpunct}{\mcitedefaultseppunct}\relax
\EndOfBibitem
\bibitem[Sun \latin{et~al.}(2022)Sun, You, Duan, Gou, Luo, Lin, Lian, Jin, Liu,
  Huang, Wang, Wee, Feng, Shen, Zhang, Chen, and Chen]{Sun2022}
Sun,~S. \latin{et~al.}  Epitaxial growth of ultraflat bismuthene with large
  topological band inversion enabled by substrate-orbital-filtering effect.
  \emph{ACS Nano} \textbf{2022}, \emph{16}, 1436--1443\relax
\mciteBstWouldAddEndPuncttrue
\mciteSetBstMidEndSepPunct{\mcitedefaultmidpunct}
{\mcitedefaultendpunct}{\mcitedefaultseppunct}\relax
\EndOfBibitem
\bibitem[Li \latin{et~al.}(2014)Li, Yu, Ye, Ge, Ou, Wu, Feng, Chen, and
  Zhang]{Li2014}
Li,~L.; Yu,~Y.; Ye,~G.~J.; Ge,~Q.; Ou,~X.; Wu,~H.; Feng,~D.; Chen,~X.~H.;
  Zhang,~Y. Black phosphorus field-effect transistors. \emph{Nature
  Nanotechnology} \textbf{2014}, \emph{9}, 372--377\relax
\mciteBstWouldAddEndPuncttrue
\mciteSetBstMidEndSepPunct{\mcitedefaultmidpunct}
{\mcitedefaultendpunct}{\mcitedefaultseppunct}\relax
\EndOfBibitem
\bibitem[Qiao \latin{et~al.}(2014)Qiao, Kong, Hu, Yang, and Ji]{Qiao2014}
Qiao,~J.; Kong,~X.; Hu,~Z.-X.; Yang,~F.; Ji,~W. High-mobility transport
  anisotropy and linear dichroism in few-layer black phosphorus. \emph{Nature
  Communications} \textbf{2014}, \emph{5}, 4475\relax
\mciteBstWouldAddEndPuncttrue
\mciteSetBstMidEndSepPunct{\mcitedefaultmidpunct}
{\mcitedefaultendpunct}{\mcitedefaultseppunct}\relax
\EndOfBibitem
\bibitem[Jin \latin{et~al.}(2019)Jin, Huang, Wang, and Liu]{Jin2019}
Jin,~K.-H.; Huang,~H.; Wang,~Z.; Liu,~F. A 2D nonsymmorphic Dirac semimetal in
  a chemically modified group-VA monolayer with a black phosphorene structure.
  \emph{Nanoscale} \textbf{2019}, \emph{11}, 7256--7262\relax
\mciteBstWouldAddEndPuncttrue
\mciteSetBstMidEndSepPunct{\mcitedefaultmidpunct}
{\mcitedefaultendpunct}{\mcitedefaultseppunct}\relax
\EndOfBibitem
\bibitem[Xiao \latin{et~al.}(2018)Xiao, Wang, Yang, Lu, Feng, and
  Zhang]{Xiao2018}
Xiao,~C.; Wang,~F.; Yang,~S.~A.; Lu,~Y.; Feng,~Y.; Zhang,~S. Elemental
  ferroelectricity and antiferroelectricity in group-V monolayer.
  \emph{Advanced Functional Materials} \textbf{2018}, \emph{28}, 1707383\relax
\mciteBstWouldAddEndPuncttrue
\mciteSetBstMidEndSepPunct{\mcitedefaultmidpunct}
{\mcitedefaultendpunct}{\mcitedefaultseppunct}\relax
\EndOfBibitem
\bibitem[Jin \latin{et~al.}(2021)Jin, Oh, Stania, Liu, and Yeom]{Jin2021}
Jin,~K.-H.; Oh,~E.; Stania,~R.; Liu,~F.; Yeom,~H.~W. Enhanced berry curvature
  dipole and persistent spin texture in the Bi(110) monolayer. \emph{Nano
  Letters} \textbf{2021}, \emph{21}, 9468--9475\relax
\mciteBstWouldAddEndPuncttrue
\mciteSetBstMidEndSepPunct{\mcitedefaultmidpunct}
{\mcitedefaultendpunct}{\mcitedefaultseppunct}\relax
\EndOfBibitem
\bibitem[Lu \latin{et~al.}(2015)Lu, Xu, Zeng, Yao, Shen, Yang, Luo, Pan, Wu,
  Das, He, Jiang, Martin, Feng, Lin, and Wang]{Lu2015}
Lu,~Y. \latin{et~al.}  Topological properties determined by atomic buckling in
  self-assembled ultrathin Bi(110). \emph{Nano Letters} \textbf{2015},
  \emph{15}, 80--87\relax
\mciteBstWouldAddEndPuncttrue
\mciteSetBstMidEndSepPunct{\mcitedefaultmidpunct}
{\mcitedefaultendpunct}{\mcitedefaultseppunct}\relax
\EndOfBibitem
\bibitem[Li \latin{et~al.}(2017)Li, Ji, Li, Hu, Cai, Zhang, and Yan]{Li2017}
Li,~S.-s.; Ji,~W.-x.; Li,~P.; Hu,~S.-j.; Cai,~L.; Zhang,~C.-w.; Yan,~S.-s.
  Tunability of the quantum spin hall effect in Bi(110) films: Effects of
  electric field and strain engineering. \emph{ACS Applied Materials \&
  Interfaces} \textbf{2017}, \emph{9}, 21515--21523\relax
\mciteBstWouldAddEndPuncttrue
\mciteSetBstMidEndSepPunct{\mcitedefaultmidpunct}
{\mcitedefaultendpunct}{\mcitedefaultseppunct}\relax
\EndOfBibitem
\bibitem[Bian \latin{et~al.}(2014)Bian, Wang, Miller, Chiang, Kowalczyk,
  Mahapatra, and Brown]{Bian2014}
Bian,~G.; Wang,~X.; Miller,~T.; Chiang,~T.-C.; Kowalczyk,~P.~J.; Mahapatra,~O.;
  Brown,~S.~A. First-principles and spectroscopic studies of Bi(110) films:
  Thickness-dependent Dirac modes and property oscillations. \emph{Phys. Rev.
  B} \textbf{2014}, \emph{90}, 195409\relax
\mciteBstWouldAddEndPuncttrue
\mciteSetBstMidEndSepPunct{\mcitedefaultmidpunct}
{\mcitedefaultendpunct}{\mcitedefaultseppunct}\relax
\EndOfBibitem
\bibitem[Wieder and Kane(2016)Wieder, and Kane]{Wieder2016}
Wieder,~B.~J.; Kane,~C.~L. Spin-orbit semimetals in the layer groups.
  \emph{Phys. Rev. B} \textbf{2016}, \emph{94}, 155108\relax
\mciteBstWouldAddEndPuncttrue
\mciteSetBstMidEndSepPunct{\mcitedefaultmidpunct}
{\mcitedefaultendpunct}{\mcitedefaultseppunct}\relax
\EndOfBibitem
\bibitem[Rodin \latin{et~al.}(2017)Rodin, Hanakata, Carvalho, Park, Campbell,
  and Castro~Neto]{Robin2017}
Rodin,~A.~S.; Hanakata,~P.~Z.; Carvalho,~A.; Park,~H.~S.; Campbell,~D.~K.;
  Castro~Neto,~A.~H. Rashba-like dispersion in buckled square lattices.
  \emph{Phys. Rev. B} \textbf{2017}, \emph{96}, 115450\relax
\mciteBstWouldAddEndPuncttrue
\mciteSetBstMidEndSepPunct{\mcitedefaultmidpunct}
{\mcitedefaultendpunct}{\mcitedefaultseppunct}\relax
\EndOfBibitem
\bibitem[Kowalczyk \latin{et~al.}(2017)Kowalczyk, Mahapatra, Le~Ster, Brown,
  Bian, Wang, and Chiang]{Kowalczyk2017}
Kowalczyk,~P.~J.; Mahapatra,~O.; Le~Ster,~M.; Brown,~S.~A.; Bian,~G.; Wang,~X.;
  Chiang,~T.-C. Single atomic layer allotrope of bismuth with rectangular
  symmetry. \emph{Phys. Rev. B} \textbf{2017}, \emph{96}, 205434\relax
\mciteBstWouldAddEndPuncttrue
\mciteSetBstMidEndSepPunct{\mcitedefaultmidpunct}
{\mcitedefaultendpunct}{\mcitedefaultseppunct}\relax
\EndOfBibitem
\bibitem[Kowalczyk \latin{et~al.}(2011)Kowalczyk, Mahapatra, McCarthy,
  Kozlowski, Klusek, and Brown]{Kowalczyk2011}
Kowalczyk,~P.~J.; Mahapatra,~O.; McCarthy,~D.; Kozlowski,~W.; Klusek,~Z.;
  Brown,~S.~A. STM and XPS investigations of bismuth islands on HOPG.
  \emph{Surface Science} \textbf{2011}, \emph{605}, 659--667\relax
\mciteBstWouldAddEndPuncttrue
\mciteSetBstMidEndSepPunct{\mcitedefaultmidpunct}
{\mcitedefaultendpunct}{\mcitedefaultseppunct}\relax
\EndOfBibitem
\bibitem[Kowalczyk \latin{et~al.}(2013)Kowalczyk, Mahapatra, Brown, Bian, Wang,
  and Chiang]{Kowalczyk2013}
Kowalczyk,~P.~J.; Mahapatra,~O.; Brown,~S.~A.; Bian,~G.; Wang,~X.;
  Chiang,~T.-C. Electronic size effects in three-dimensional nanostructures.
  \emph{Nano Letters} \textbf{2013}, \emph{13}, 43--47\relax
\mciteBstWouldAddEndPuncttrue
\mciteSetBstMidEndSepPunct{\mcitedefaultmidpunct}
{\mcitedefaultendpunct}{\mcitedefaultseppunct}\relax
\EndOfBibitem
\bibitem[Nagase \latin{et~al.}(2018)Nagase, Kokubo, Yamazaki, Nakatsuji, and
  Hirayama]{Nagase2018}
Nagase,~K.; Kokubo,~I.; Yamazaki,~S.; Nakatsuji,~K.; Hirayama,~H. Structure and
  growth of Bi(110) islands on Si(111)$\sqrt{3}$$\times$$\sqrt{3}$-\textit{B}
  substrates. \emph{Phys. Rev. B} \textbf{2018}, \emph{97}, 195418\relax
\mciteBstWouldAddEndPuncttrue
\mciteSetBstMidEndSepPunct{\mcitedefaultmidpunct}
{\mcitedefaultendpunct}{\mcitedefaultseppunct}\relax
\EndOfBibitem
\bibitem[Wang \latin{et~al.}(2022)Wang, Sun, Du, Ma, and Wang]{Wang2022}
Wang,~J.; Sun,~X.; Du,~H.; Ma,~C.; Wang,~B. Electronic and topological
  properties of Bi(110) ultrathin films grown on a Cu(111) substrate.
  \emph{Phys. Rev. B} \textbf{2022}, \emph{105}, 115407\relax
\mciteBstWouldAddEndPuncttrue
\mciteSetBstMidEndSepPunct{\mcitedefaultmidpunct}
{\mcitedefaultendpunct}{\mcitedefaultseppunct}\relax
\EndOfBibitem
\bibitem[Pan \latin{et~al.}(2011)Pan, Liu, Ming, Wang, and Xiao]{Pan2011}
Pan,~S.; Liu,~Q.; Ming,~F.; Wang,~K.; Xiao,~X. Interface effects on the quantum
  well states of Pb thin films. \emph{Journal of Physics: Condensed Matter}
  \textbf{2011}, \emph{23}, 485001\relax
\mciteBstWouldAddEndPuncttrue
\mciteSetBstMidEndSepPunct{\mcitedefaultmidpunct}
{\mcitedefaultendpunct}{\mcitedefaultseppunct}\relax
\EndOfBibitem
\bibitem[Özer \latin{et~al.}(2007)Özer, Jia, Zhang, Thompson, and
  Weitering]{Mustafa2007}
Özer,~M.~M.; Jia,~Y.; Zhang,~Z.; Thompson,~J.~R.; Weitering,~H.~H. Tuning the
  quantum stability and superconductivity of ultrathin metal alloys.
  \emph{Science} \textbf{2007}, \emph{316}, 1594--1597\relax
\mciteBstWouldAddEndPuncttrue
\mciteSetBstMidEndSepPunct{\mcitedefaultmidpunct}
{\mcitedefaultendpunct}{\mcitedefaultseppunct}\relax
\EndOfBibitem
\bibitem[Ming-Yang \latin{et~al.}(2021)Ming-Yang, Ju-Feng, Hong-Jian, Chuan-Xu,
  and Bing]{Ming2021}
Ming-Yang,~T.; Ju-Feng,~W.; Hong-Jian,~D.; Chuan-Xu,~M.; Bing,~W.
  Characterization of structure and superconducting properties of
  low-temperature phase of Pb-Bi alloy films. \emph{ACTA PHYSICA SINICA}
  \textbf{2021}, \emph{70}, 170703\relax
\mciteBstWouldAddEndPuncttrue
\mciteSetBstMidEndSepPunct{\mcitedefaultmidpunct}
{\mcitedefaultendpunct}{\mcitedefaultseppunct}\relax
\EndOfBibitem
\bibitem[Peng \latin{et~al.}(2019)Peng, Qiao, Xian, Pan, Ji, Zhang, and
  Fu]{Peng2019}
Peng,~L.; Qiao,~J.; Xian,~J.-J.; Pan,~Y.; Ji,~W.; Zhang,~W.; Fu,~Y.-S. Unusual
  electronic states and superconducting proximity effect of Bi films modulated
  by a NbSe$_{2}$ substrate. \emph{ACS Nano} \textbf{2019}, \emph{13},
  1885--1892\relax
\mciteBstWouldAddEndPuncttrue
\mciteSetBstMidEndSepPunct{\mcitedefaultmidpunct}
{\mcitedefaultendpunct}{\mcitedefaultseppunct}\relax
\EndOfBibitem
\bibitem[Sun \latin{et~al.}(2012)Sun, Huang, Wong, Gao, Feng, and Wee]{Sun2012}
Sun,~J.-T.; Huang,~H.; Wong,~S.~L.; Gao,~H.-J.; Feng,~Y.~P.; Wee,~A. T.~S.
  Energy-gap opening in a Bi(110) nanoribbon induced by edge reconstruction.
  \emph{Phys. Rev. Lett.} \textbf{2012}, \emph{109}, 246804\relax
\mciteBstWouldAddEndPuncttrue
\mciteSetBstMidEndSepPunct{\mcitedefaultmidpunct}
{\mcitedefaultendpunct}{\mcitedefaultseppunct}\relax
\EndOfBibitem
\bibitem[Marchenko \latin{et~al.}(2012)Marchenko, Varykhalov, Scholz,
  Bihlmayer, Rashba, Rybkin, Shikin, and Rader]{Marchenko2012}
Marchenko,~D.; Varykhalov,~A.; Scholz,~M.~R.; Bihlmayer,~G.; Rashba,~E.~I.;
  Rybkin,~A.; Shikin,~A.~M.; Rader,~O. Giant Rashba splitting in graphene due
  to hybridization with gold. \emph{Nature Communications} \textbf{2012},
  \emph{3}, 1232\relax
\mciteBstWouldAddEndPuncttrue
\mciteSetBstMidEndSepPunct{\mcitedefaultmidpunct}
{\mcitedefaultendpunct}{\mcitedefaultseppunct}\relax
\EndOfBibitem
\bibitem[Liu \latin{et~al.}(2011)Liu, Wu, Ming, Zhang, Wang, Wang, and
  Xiao]{Liu2011_1}
Liu,~J.; Wu,~X.; Ming,~F.; Zhang,~X.; Wang,~K.; Wang,~B.; Xiao,~X.
  Size-dependent superconducting state of individual nanosized Pb islands grown
  on Si(111) by tunneling spectroscopy. \emph{Journal of Physics: Condensed
  Matter} \textbf{2011}, \emph{23}, 265007\relax
\mciteBstWouldAddEndPuncttrue
\mciteSetBstMidEndSepPunct{\mcitedefaultmidpunct}
{\mcitedefaultendpunct}{\mcitedefaultseppunct}\relax
\EndOfBibitem
\bibitem[Eom \latin{et~al.}(2006)Eom, Qin, Chou, and Shih]{Eom2006}
Eom,~D.; Qin,~S.; Chou,~M.-Y.; Shih,~C.~K. Persistent superconductivity in
  ultrathin Pb films: A scanning tunneling spectroscopy study. \emph{Phys. Rev.
  Lett.} \textbf{2006}, \emph{96}, 027005\relax
\mciteBstWouldAddEndPuncttrue
\mciteSetBstMidEndSepPunct{\mcitedefaultmidpunct}
{\mcitedefaultendpunct}{\mcitedefaultseppunct}\relax
\EndOfBibitem
\bibitem[Mans \latin{et~al.}(2002)Mans, Dil, Ettema, and Weitering]{Mans2002}
Mans,~A.; Dil,~J.~H.; Ettema,~A. R. H.~F.; Weitering,~H.~H. Quantum electronic
  stability and spectroscopy of ultrathin Pb films on Si(111)7$\times$7.
  \emph{Phys. Rev. B} \textbf{2002}, \emph{66}, 195410\relax
\mciteBstWouldAddEndPuncttrue
\mciteSetBstMidEndSepPunct{\mcitedefaultmidpunct}
{\mcitedefaultendpunct}{\mcitedefaultseppunct}\relax
\EndOfBibitem
\bibitem[Wei and Chou(2002)Wei, and Chou]{Wei2002}
Wei,~C.~M.; Chou,~M.~Y. Theory of quantum size effects in thin Pb(111) films.
  \emph{Phys. Rev. B} \textbf{2002}, \emph{66}, 233408\relax
\mciteBstWouldAddEndPuncttrue
\mciteSetBstMidEndSepPunct{\mcitedefaultmidpunct}
{\mcitedefaultendpunct}{\mcitedefaultseppunct}\relax
\EndOfBibitem
\bibitem[Hong \latin{et~al.}(2009)Hong, Brun, Patthey, Sklyadneva, Zubizarreta,
  Heid, Silkin, Echenique, Bohnen, Chulkov, and Schneider]{Hong2009}
Hong,~I.-P.; Brun,~C.; Patthey,~F. m.~c.; Sklyadneva,~I.~Y.; Zubizarreta,~X.;
  Heid,~R.; Silkin,~V.~M.; Echenique,~P.~M.; Bohnen,~K.~P.; Chulkov,~E.~V.;
  Schneider,~W.-D. Decay mechanisms of excited electrons in quantum-well states
  of ultrathin Pb islands grown on Si(111): Scanning tunneling spectroscopy and
  theory. \emph{Phys. Rev. B} \textbf{2009}, \emph{80}, 081409\relax
\mciteBstWouldAddEndPuncttrue
\mciteSetBstMidEndSepPunct{\mcitedefaultmidpunct}
{\mcitedefaultendpunct}{\mcitedefaultseppunct}\relax
\EndOfBibitem
\bibitem[Vyalikh \latin{et~al.}(2003)Vyalikh, Weschke, Dedkov, Kaindl, Shikin,
  and Adamchuk]{Vyalikh2003}
Vyalikh,~D.; Weschke,~E.; Dedkov,~Y.; Kaindl,~G.; Shikin,~A.; Adamchuk,~V.
  Quantum-well states in bilayers of Ag and Au on W(110). \emph{Surface
  Science} \textbf{2003}, \emph{540}, L638--L642\relax
\mciteBstWouldAddEndPuncttrue
\mciteSetBstMidEndSepPunct{\mcitedefaultmidpunct}
{\mcitedefaultendpunct}{\mcitedefaultseppunct}\relax
\EndOfBibitem
\bibitem[Pan \latin{et~al.}(2020)Pan, Lee, Shih, and Chou]{Pan2020}
Pan,~C.-R.; Lee,~W.; Shih,~C.-K.; Chou,~M.~Y. Coherently coupled quantum-well
  states in bimetallic Pb/Ag thin films. \emph{Phys. Rev. B} \textbf{2020},
  \emph{102}, 115428\relax
\mciteBstWouldAddEndPuncttrue
\mciteSetBstMidEndSepPunct{\mcitedefaultmidpunct}
{\mcitedefaultendpunct}{\mcitedefaultseppunct}\relax
\EndOfBibitem
\bibitem[Chiang(2000)]{Chiang2000}
Chiang,~T.-C. Photoemission studies of quantum well states in thin films.
  \emph{Surface Science Reports} \textbf{2000}, \emph{39}, 181--235\relax
\mciteBstWouldAddEndPuncttrue
\mciteSetBstMidEndSepPunct{\mcitedefaultmidpunct}
{\mcitedefaultendpunct}{\mcitedefaultseppunct}\relax
\EndOfBibitem
\bibitem[Schnyder \latin{et~al.}(2008)Schnyder, Ryu, Furusaki, and
  Ludwig]{Schnyder2008}
Schnyder,~A.~P.; Ryu,~S.; Furusaki,~A.; Ludwig,~A. W.~W. Classification of
  topological insulators and superconductors in three spatial dimensions.
  \emph{Phys. Rev. B} \textbf{2008}, \emph{78}, 195125\relax
\mciteBstWouldAddEndPuncttrue
\mciteSetBstMidEndSepPunct{\mcitedefaultmidpunct}
{\mcitedefaultendpunct}{\mcitedefaultseppunct}\relax
\EndOfBibitem
\bibitem[Jin \latin{et~al.}(2019)Jin, Huang, Mei, Liu, Lim, and Liu]{Jin2019_1}
Jin,~K.-H.; Huang,~H.; Mei,~J.-W.; Liu,~Z.; Lim,~L.-K.; Liu,~F. Topological
  superconducting phase in high-$T_{c}$ superconductor MgB$_{2}$ with
  Dirac-nodal-line fermions. \emph{npj Computational Materials} \textbf{2019},
  \emph{5}, 57\relax
\mciteBstWouldAddEndPuncttrue
\mciteSetBstMidEndSepPunct{\mcitedefaultmidpunct}
{\mcitedefaultendpunct}{\mcitedefaultseppunct}\relax
\EndOfBibitem
\bibitem[Zhang \latin{et~al.}(2021)Zhang, Jin, Mao, Zhao, Liu, and
  Liu]{Zhang2021}
Zhang,~X.; Jin,~K.-H.; Mao,~J.; Zhao,~M.; Liu,~Z.; Liu,~F. Prediction of
  intrinsic topological superconductivity in Mn-doped GeTe monolayer from
  first-principles. \emph{npj Computational Materials} \textbf{2021}, \emph{7},
  44\relax
\mciteBstWouldAddEndPuncttrue
\mciteSetBstMidEndSepPunct{\mcitedefaultmidpunct}
{\mcitedefaultendpunct}{\mcitedefaultseppunct}\relax
\EndOfBibitem
\bibitem[Verd\'un and Drew(1976)Verd\'un, and Drew]{Verdun1976}
Verd\'un,~H.~R.; Drew,~H.~D. Far-infrared magnetospectroscopy of the hole
  pocket in bismuth. I. Band-structure effects. \emph{Phys. Rev. B}
  \textbf{1976}, \emph{14}, 1370--1394\relax
\mciteBstWouldAddEndPuncttrue
\mciteSetBstMidEndSepPunct{\mcitedefaultmidpunct}
{\mcitedefaultendpunct}{\mcitedefaultseppunct}\relax
\EndOfBibitem
\bibitem[Hsu and Valles(1993)Hsu, and Valles]{Hsu1993}
Hsu,~S.-Y.; Valles,~J.~M. Perpendicular upper critical field of granular Pb
  films near the superconductor-to-insulator transition. \emph{Phys. Rev. B}
  \textbf{1993}, \emph{47}, 14334--14337\relax
\mciteBstWouldAddEndPuncttrue
\mciteSetBstMidEndSepPunct{\mcitedefaultmidpunct}
{\mcitedefaultendpunct}{\mcitedefaultseppunct}\relax
\EndOfBibitem
\bibitem[Liu \latin{et~al.}(2018)Liu, Wang, Zhang, Liu, Liu, Zhou, Wang, Wang,
  Liu, Xi, Tian, Liu, Feng, Xie, and Wang]{Liu2018}
Liu,~Y.; Wang,~Z.; Zhang,~X.; Liu,~C.; Liu,~Y.; Zhou,~Z.; Wang,~J.; Wang,~Q.;
  Liu,~Y.; Xi,~C.; Tian,~M.; Liu,~H.; Feng,~J.; Xie,~X.~C.; Wang,~J.
  Interface-Induced Zeeman-Protected Superconductivity in Ultrathin Crystalline
  Lead Films. \emph{Phys. Rev. X} \textbf{2018}, \emph{8}, 021002\relax
\mciteBstWouldAddEndPuncttrue
\mciteSetBstMidEndSepPunct{\mcitedefaultmidpunct}
{\mcitedefaultendpunct}{\mcitedefaultseppunct}\relax
\EndOfBibitem
\bibitem[Kresse and Furthm\"uller(1996)Kresse, and Furthm\"uller]{Kresse1996}
Kresse,~G.; Furthm\"uller,~J. Efficient iterative schemes for \textit{ab
  initio} total-energy calculations using a plane-wave basis set. \emph{Phys.
  Rev. B} \textbf{1996}, \emph{54}, 11169--11186\relax
\mciteBstWouldAddEndPuncttrue
\mciteSetBstMidEndSepPunct{\mcitedefaultmidpunct}
{\mcitedefaultendpunct}{\mcitedefaultseppunct}\relax
\EndOfBibitem
\bibitem[Perdew \latin{et~al.}(1996)Perdew, Burke, and Ernzerhof]{Perdew1996}
Perdew,~J.~P.; Burke,~K.; Ernzerhof,~M. Generalized gradient approximation made
  simple. \emph{Phys. Rev. Lett.} \textbf{1996}, \emph{77}, 3865--3868\relax
\mciteBstWouldAddEndPuncttrue
\mciteSetBstMidEndSepPunct{\mcitedefaultmidpunct}
{\mcitedefaultendpunct}{\mcitedefaultseppunct}\relax
\EndOfBibitem
\end{mcitethebibliography}

\end{document}